\documentclass{kluwer}    

\usepackage{klups}
\usepackage[dvips]{graphicx}

\newdisplay{guess}{Conjecture}

\begin{document}                                                                                   
\begin{article}
\begin{opening}         

\title{Evolution of Magnetic fields in Bok globules?} 

\author{Sebastian \surname{Wolf}}  
\runningauthor{Wolf, Launhardt, \& Henning}
\runningtitle{Evolution of Magnetic fields in Bok globules?}
\institute{California Institute of Technology, 1201 East California Blvd, 
  Mail code 105-24, Pasadena, CA 91125, USA; swolf@astro.caltech.edu}

\author{Ralf \surname{Launhardt}}  
\author{Thomas \surname{Henning}}  
\institute{Max-Planck-Institut f\"ur Astronomie, K\"onigstuhl 17, D-69117 Heidelberg, Germany; 
  rl@mpia-hd.mpg.de, henning@mpia-hd.mpg.de}

\begin{abstract}
    We study the influence and structure of the magnetic field in the early phases of low-mass star formation
    using polarization maps of Bok globules at a wavelength of 850\,$\mu$m, obtained with 
    the Submillimeter Common-User Bolometer Array (SCUBA) at the James Clerk Maxwell Telescope (JCMT). 
    We discuss observations of the following sources: 
    CB~26 - a small globule with a nearly dispersed dense core and a young and large circumstellar disk,
    CB~54 - a large globule with a massive dense core and a deeply embedded young stellar cluster, and
    B~335, CB~230, and CB~244 - three nearby, relatively isolated small globules with low-mass protostellar cores.

    We find strongly aligned polarization vectors in the case of
    CB~26, B~335, and CB~230, while the vector orientations in the case of CB~54 and CB~244
    are more or less randomly distributed. The degree of polarization,
    amounting to several percent, was found to decrease toward the center in each source.
    Assuming dichroic emission by aligned non-spherical grains as the polarization mechanism, where the magnetic field
    plays a role in the alignment process, we derive magnetic field strengths and structures
    from the observed polarization patterns.

    We compare the magnetic field topology with the morphology and outflow directions of the globules.
    In the Class~0 sources B~335, CB~230, and CB~244, the magnetic field is
    oriented almost perpendicular to the ouflow direction.
    In contrast, the inclination between outflow axis and magnetic field direction
    is much more moderate (36$^{\rm o}$) in the more evolved Class~I source CB26.
\end{abstract}


\end{opening}

\section{Introduction}\label{intro}

Bok globules are small, relatively isolated, simply-structured molecular clouds
and therefore excellent objects to study the earliest processes of star formation,
in particular the interplay between protostellar collapse, fragmentation, and magnetic fields.
Signs of star formation in many Bok globules include bipolar molecular outflows
(e.g., Yun \& Clemens 1994a), infrared colors, and sub-millimeter
properties that are consistent with Class~0 protostars or embedded
Class~I sources (Yun \& Clemens~1994b, Launhardt \& Henning~1997).

  To study one of the key parameters of the star formation process
-- namely the magnetic field -- submillimeter polarization measurements represent a powerful technique.
Assuming emission by aligned nonspherical grains as the dominating
polarization mechanism, where the magnetic field plays a role in the alignment process,
magnetic field strengths and structures can be derived from the submillimeter
polarization pattern (e.g.\ Wolf et al.~2003, Henning et al.~2001).

\section{Observations and Data reduction} \label{obs}

We observed the Bok globules CB\,26, CB\,54, B\,335, CB230, and CB\,244 for which 
brief descriptions are given in the following:
{\bf CB\,26} is a small, slightly cometary-shaped Bok globule at $D \sim 140$\,pc.
The globule contains a nearly dispersed dense core and a young and large, 
perhaps early protoplanetary, disk (Launhardt \& Sargent 2001).
{\bf CB\,54} is a large Bok globule associated with the molecular cloud BBW\,4 
at $D \sim 1.1$\,kpc (Brand \& Blitz 1993) and the reflection nebula 
LBN\,1042. The globule contains a massive dense core 
of $M_{\rm H} \sim 100$\,M$_{\rm sun}$, which is associated with a bipolar 
molecular outflow (Launhardt \& Henning 1997; Yun \& Clemens 1994a).
At near-infrared wavelengths, a small, deeply embedded young stellar cluster 
becomes visible.
{\bf B\,335} is an isolated, nearly spherical Bok globule at a distance of 
$\sim$\,250\,pc (Tomita et al.~1979, Frerking et al.~1987). 
The deeply embedded Class~0 protostar drives a collimated bipolar outflow with a dynamical age of $\sim
3\times 10^4$\,yr  (e.g.\ Chandler et al.~1990, Chandler \& Sargent~1993).
{\bf CB230} is a small, bright-rimmed Bok globule in a distance of 400$\pm$100\,pc
associated with the Cepheus Flare molecular cloud complex.
The globule contains a protostellar double core with 10'' separation (east-west). 
The western source with the more massive accretion disk and envelope drives 
a large-scale collimated molecular outflow of dynamical age $\sim 2\times 10^4$\,yr 
(Yun \& Clemens~1994a). The eastern source drives a weaker outflow that is probably 
not aligned with the large-scale outflow (Launhardt et al., in prep.).
{\bf CB\,244} is an isolated Bok globule located at a distance of 180\,pc and 
probably associated with the Lindblad ring (Kun~1998). 
It contains two dense cores separated by $\sim$\,\,90''. The more prominent 
south-eastern core is associated with a bipolar molecular outflow 
with a dynamical age of $\sim 10^4$\,yr (Yun \& Clemens 1994a; Launhardt et al.~1997).

The observations were performed at the 15-m James Clerk Maxwell Telescope (JCMT)
between March 1 and 6, 2000 and between September 10 and 14, 2001.
The effective beam size (HPBW) is $\sim$\,14.7'' at 850\,$\mu$m.
Polarimetry was conducted with the Submillimeter Common-User Bolometer Array 
(SCUBA; Holland et al.~1999) and its polarimeter, SCU-POL, using the 350-850\,$\mu$m achromatic half-waveplate.
Flat-fielding, extinction correction, sky-noise and instrumental polarization removal have been performed with
the data reduction package SURF (Jenness \& Lightfoot~1998).
The polarization was computed using the POLPACK data reduction package (Berry \& Gledhill 1999).

\begin{figure*}
  \resizebox{\hsize}{!}{\includegraphics{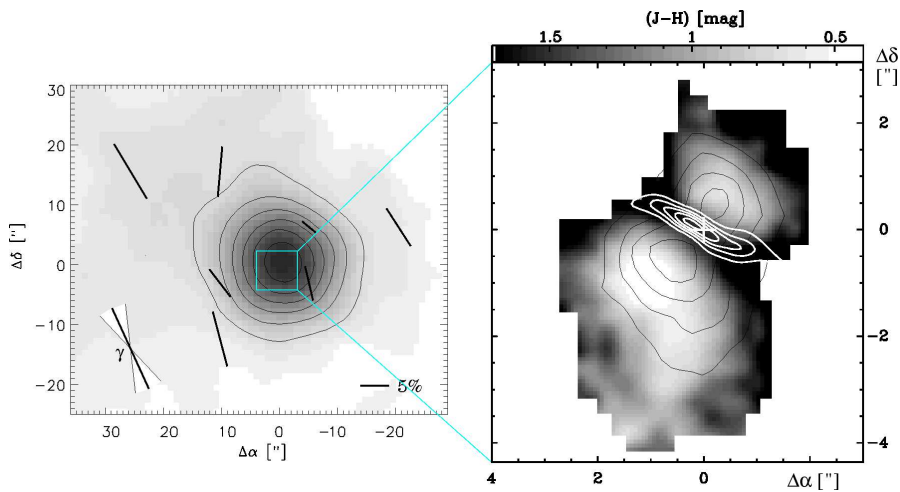}}
  \caption[]{
    {\bf Left:} SCUBA intensity map (850\,$\mu$m) of the Bok globule CB~26 with overlayed
    iso-intensity contour lines (contour levels: 
    0.15\%, 0.30\%, $\ldots$, 0.90\% of the maximum intensity) and polarization pattern.
    Furthermore, the mean direction of the polarization $\bar{\gamma}(\pm\sigma_{\bar{\gamma}})$ is shown.
    {\bf Right:}
    Central part of the Bok globule CB~26 (from Launhardt \& Sargent~2001).
    $J-H$ color map of the bipolar near-infrared reflection nebula (black contour lines: $K$ band emission).
    The white contour lines show the proto-planetary disk discovered by Launhardt \& Sargent~(2001;
    contour levels: 4, 11, $\ldots$, 32 mJy arcsec$^{-2}$; obtained with OVRO at 1.3\,mm - beam width: 
    0.58''$\times$0.39''). {\em (from Wolf et al.~2003)}
  }
  \label{of-cb26}
\end{figure*}

\begin{figure*}
  \resizebox{\hsize}{!}{\includegraphics{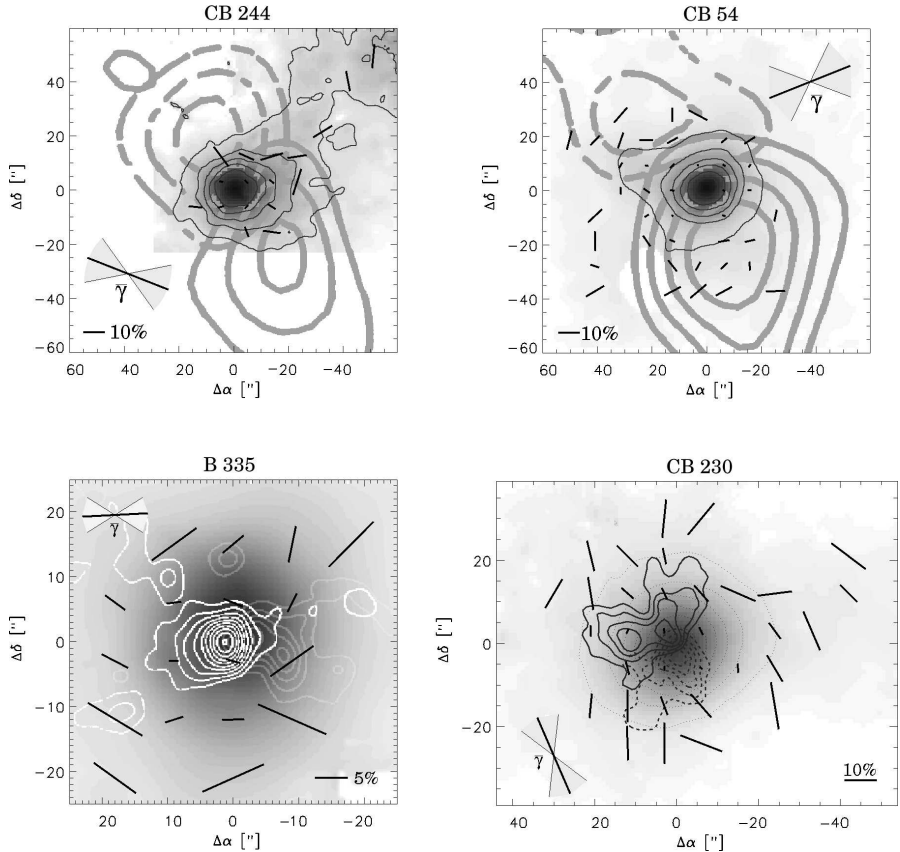}}
  \caption[]{SCUBA intensity maps (850\,$\mu$m) with overlayed iso-intensity contour lines
    (contour levels: 0.15\%, 0.30\%, $\ldots$, 0.90\% of the maximum intensity) and
    polarization patterns of the Bok globules CB~244, CB~54, B~335, and CB~230.
    Furthermore, the solid/dashed contour lines represent the spatially well-separated
    blue-shifted/red-shifted outflow lobe seen in either the
    $^{12}$CO(1-0) or $^{13}$CO(1-0) spectral channel maps of
    {\bf CB~244} (blue/red-shifted outflow contour lines begin with 0.5\,K\,km\,s$^{-1}$/0.6\,K\,km\,s$^{-1}$
    and are stepped by 0.15\,K\,km\,s$^{-1}$/0.2\,K\,km\,s$^{-1}$; from Yun \& Clemens~1994a),    
    {\bf CB~54} (blue/red-shifted outflow contour lines begin with 1.7\,K\,km\,s$^{-1}$/0.45\,K\,km\,s$^{-1}$
    and are stepped by 0.3\,K\,km\,s$^{-1}$/0.15\,K\,km\,s$^{-1}$; from Yun \& Clemens~1994a),
    {\bf B~335} (contours are spaced at $2\sigma$ intervals of 200\,mJy\,beam$^{-1}$; from Chandler \& Sargent~1993),
    and
    {\bf CB~230} (blue lobe: $v_{\rm LSR}=1.6\ldots$2.8\,km\,s$^{-1}$ - solid contours,
    red  lobe: $v_{\rm LSR}=3.0\ldots$4.2\,km\,s$^{-1}$ - dashed contours.
    The step width amounts to 0.3\,km\,s$^{-1}$ for both lobes; from Launhardt~2001)
    are overlayed. {\em (from Wolf et al.~2003)}
    }
  \label{of-b335}
\end{figure*}

We restrict the polarization analysis to the regions in which the total flux density 
per beam is above 5 times the rms in the maps (measured outside the central sources). 
We do not use polarization vectors derived at positions 
where the scatter of the total flux density measurements between 
the jiggle cycles was larger than 20\% of the average total flux density at that point and
exclude polarization vectors with $P_{\rm l} / \sigma(P_{\rm l}) < 3$, \
where $\sigma(P_{\rm l})$ is the standard deviation of the polarization degree.
The resulting polarization maps are shown in Fig.~\ref{of-cb26} and \ref{of-b335}.

\section{Polarization versus Intensity}\label{plvsi}

\begin{figure*}
  \resizebox{0.8\hsize}{!}{\includegraphics{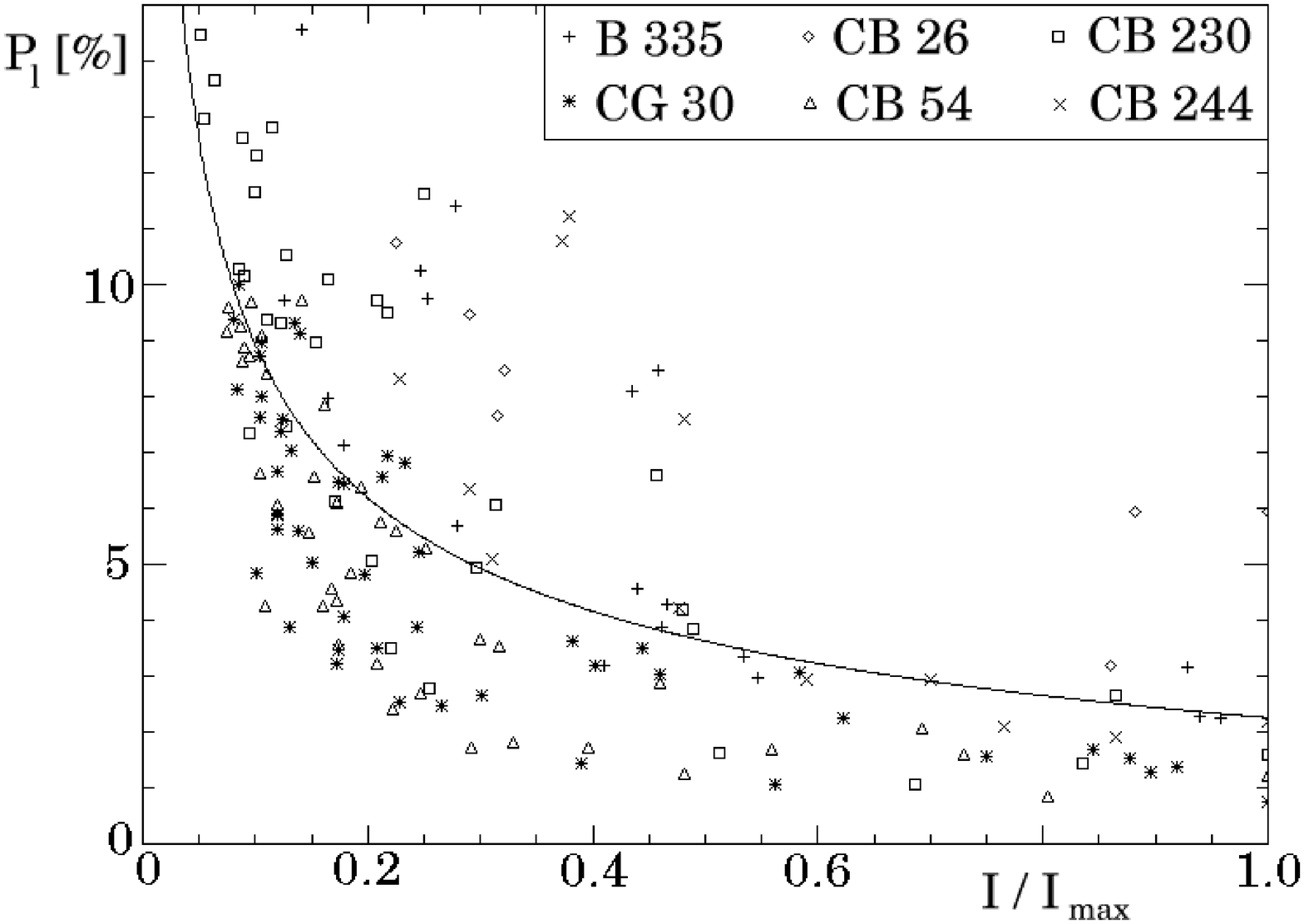}}
  \caption[]{Scatter diagram showing the distribution of the linear polarization degree $P_{\rm l}$ 
    vs.\ intensity $I$ across the Bok globules.
    Additionally, data obtained for the Bok globule CG~30 have been taken into account.
    The solid line represents the best fit of the function
    of the form $P_{\rm l} = a_0 + a_1 \left(\frac{I}{{\rm max}(I)}\right)^{a_2}$,
    where $P_{\rm l}$ is the degree of linear polarization, $I$ is the measured intensity,
    and $a_0$, $a_1$, and $a_2$ are constant quantities.
  }
  \label{scatt}
\end{figure*}

As found in previous polarization measurements towards other star-forming cores
(see, e.g., Matthews \& Wilson~2002, Houde et al.~2002, Glenn et al.~1999),
the degree of polarization decreases towards regions of increasing intensity 
(see Fig.~\ref{scatt}).
Possible explanations for this behavior are:
(a) increased densities in the cores, which cause an increase of
    the collisional disalignment rate of the grains towards the centers of the cores,
(b) insufficient resolution of the field structure associated with the core collapse, and
(c) spherical grain growth in the denser regions which would result in unpolarized 
    re-emission by the dust (Weintraub et al.~2000).
Because of the large density contrast between the unresolved central condensations
and the envelopes, the increasingly poor grain alignment as density
increases is expected to be the main effect. 

Furthermore, the observations of B~335 and CB~230 not only show a decrease of the polarization
degree, but also that the orientation of the polarization vectors 
does not change remarkably towards the dense cores. This supports the hypothesis that despite the
decrease of the grain alignment rate, the magnetic field structure in the cores of these objects
is not seriously disturbed, and thus still represents the primordial field.
In the case of CB~244 and CB~54 on the other hand, the orientation of the
polarization vectors and therefore the structure of the magnetic field in the 
cores is irregular and we assume that it can no longer be accounted
for being representative for the primordial field.

\section{Magnetic field strength and structure} \label{magfield}

Polarized thermal emission by aligned non-spherical
grains is the main source of polarized submillimeter radiation 
in Bok globules (e.g., Weintraub et al.~2000, Greaves et al.~1999).
Irrespective of the alignment mechanism, charged rotating interstellar grains would have a
substantial magnetic moment, leading to a rapid precession of the grain angular
momentum around the magnetic field direction (Draine \& Weingartner 1997). 
This implies a net alignment of the grains with the magnetic field.
Based on the work by Chandrasekhar \& Fermi (1953), the dispersion of 
polarization position angles is thought to be inversely proportional 
to the magnetic field strength
\begin{equation}\label{eqb}
B = \sqrt{\frac{4\,\pi}{3} \, \rho_{\rm Gas}} 
\cdot \frac{v_{\rm turb}}{\sigma_{\bar \gamma}}\ .
\end{equation}
Here, $B$ is the magnetic field strength in units of G,
$\rho_{\rm Gas}$ is the gas density (in units of g\,$\rm cm^{-3}$),
$v_{\rm turb}$ the rms turbulence velocity (in units of cm\,$\rm s^{-1}$), and
$\sigma_{\bar{\gamma}}$ the standard deviation to the mean orientation 
angle $\bar{\gamma}$
of the polarization vectors (in units of radians).
The resulting magnetic field strengths are in the range of $\approx 100 - 260 \mu$G (see Tabl.~\ref{tablegauss}).
\begin{table}
\begin{tabular}{lcccc}
{Object}                    &
{$\rho_{\rm Gas}$}          & 
{$v_{\rm turb}$}            & 
{$\sigma_{\bar{\gamma}}$}   & 
{B}                         \\
{}                          & 
{[g\,${\rm cm^{-3}}]$}      & 
{[km\,${\rm s^{-1}}]$}      &
{[$^{\rm o}$]}                & 
{[$\mu$G]}                  \\
B\,335  (CB 199)           & 8.6E-18 &  0.14$\pm$0.02$^{\rm a}$ &  $35.8^{+14.6}_{-9.1}$   &  134$^{+46}_{-39}$  \\
CB\,230 (L 1177)           & 3.6E-18 &  0.29$\pm$0.04$^{\rm b}$ &  $29.8^{+8.8}_{-6.1}$    &  218$^{+56}_{-50}$  \\
CB\,244 (L 1262, SE core)  & 8.0E-18 &  $\approx$0.29$^{\rm c}$ &  $33.1^{+19.2}_{-10.4}$  &  257$^{+111}_{-91}$ \\
CB\,26  (L 1439)           & 8.6E-19 &           0.25$^{\rm d}$ &  $18.9^{+16.7}_{-7.3}$   &  144$^{+91}_{-68}$  \\
CB\,54                     & 3.4E-19 &           0.65$^{\rm c}$ &  $42.7^{+11.1}_{-8.0}$   &  104$^{+24}_{-21}$  \\
\end{tabular}
\caption{Masses, gas densities, polarization, and magnetic field
strengths of the envelopes.
(a) Frerking et al.~1987.
(b) Wang et al.~1995; Launhardt et al.~1996.
(c) Wang et al.~1995.
(d) RMS turbulence velocity of a large sample of nearby star-forming Bok globules 
derived from C$^{18}$O (J=2-1).
}
\label{tablegauss}
\end{table}

The comparison between the magnetic field structure and morphologic features of the globules,
such as their elongation and outflow direction, reveals following correlations:
First,  the dense cores in B~335 and CB~230 are slightly elongated whereby the major axes
        are oriented almost perpendicular to the direction of the bipolar molecular outflows.
Second, in the Class~0 source harbouring Bok globules B~335, CB~230, and CB~244, 
        the outflows are oriented almost perpendicular to
	the mean direction of the magnetic field as seen in projection on the plane of the sky.
        In contrast, the directions of the magnetic field and ouflow are aligned more parallel
        in the massive globule core CB~54 and in CB~26 with an embedded more evolved Class~I source.
        We want to remark that the statistical significance of the latter is low
	because of the large scattering of polarization directions measured in CB~54
	and the small number of polarization vectors and low spatial 
	resolution in the case of in CB~26.

\section{Conclusions} \label{concl}

We presented 850\,$\mu$m polarimetric observations of dense envelopes around the very high-density 
protostellar condensations in five selected Bok globules.
In each object we found polarization degrees of several percent which decrease towards the cores of the globules.
Since the functional dependence of this behavior between the intensity and polarization
is very similar for each globule, the optical properties of the grains do not seem to play a key role
for the observed polarization decrease, but merely the coupling of the magnetic
field to the grains.

Using the formalism by Chandrasekhar \& Fermi~(1953) we derive
an estimate of the mean magnetic field strengths in these globules 
on the order of several hundred $\mu$G.
These magnetic field strengths are well above those of the interstellar medium (see, e.g., Myers et al.\ 1995)
but comparable to typical magnetic field strengths found in molecular clouds,
pre-protostellar cores, and other star-forming regions.

We find the direction of the polarization vectors/magnetic field, 
the elongation of the Bok globules, and
the orientation of the outflows (in the case of CB~26: potential outflow direction along the lobes) 
to be related to one another.
In particular, we find the mean magnetic field direction of the Bok globules with embedded
Class~0 sources -- B~335, CB~230, and CB~244 -- to be oriented almost perpendicular to the outflow axis.
In contrast to this, the magnetic field is nearly aligned with the potential outflow
direction of CB~26, harbouring a more evolved Class~I source.

Our findings suggests that the mean magnetic field direction in a protostellar envelope
changes from initially perpendicular to the outflow axis to a more moderate inclination
at the end of the main accretion/ouflow phase.
However, a larger sample of protostellar sources in different
evolutionary stages has to be investigated to confirm this preliminary conclusion.


\acknowledgements
S.W.\ acknowledges support through the HST Grant GO\,9160, 
and through the NASA grant NAG5-11645.
JCMT is operated by the Joint Astronomical Centre on behalf of the UK Particle
Physics and Astronomy Research Council.



\end{article}
\end{document}